\documentclass{lni}

\usepackage[style=numeric,backend=biber]{biblatex}

\usepackage{xcolor}
\usepackage{comment}
\usepackage{todonotes}
\usepackage{listings}
 \lstdefinestyle{codeLine}{
    language=Java,
 	breaklines=true,
 	tabsize=2,
	belowskip=-0.7ex,
	aboveskip=2ex,
	framexleftmargin=5mm, 
    frame=lrtb,
    framesep=1pt,
    rulesepcolor=\color{black},
    basicstyle=\ttfamily\scriptsize,
    numbers=left, 
    numberstyle=\tiny, 
    stepnumber=1, 
    numbersep=5pt
 }
\excludecomment{skipit}

\newcommand{\eUCRITE}[1]{eUCRITE}

\excludecomment{full}

\newbool{short}
\setboolean{short}{true}

\begin{document}

\title[PQC-Integration in Softwareprodukte]{Zur Integration von Post-Quantum Verfahren in bestehende Softwareprodukte}

\author[Zeier, Wiesmaier, Heinemann]{Alexander Zeier\footnote{Hochschule Darmstadt, FB Informatik, Haardtring 100, 64295 Darmstadt \email{<vorname>.<nachname>@h-da.de}} 
\and
Alexander Wiesmaier\footnotemark[1]
%{Hochschule Darmstadt, FB Informatik, Haardtring 100, 64295 Darmstadt
%\email{alexander.wiesmaier@h-da.de}}
\and
Andreas Heinemann\footnotemark[1]
%{Hochschule Darmstadt, FB Informatik, Haardtring 100, 64295 Darmstadt
%\email{andreas.heinemann@h-da.de}}
}

\startpage{1} 
\editor{} 
\booktitle{Beitrag für den 17. Deutschen IT-Sicherheitskongress, 2021} 
\year{2021}

\maketitle

\section*{Zusammenfassung}
% Setting the scene
Aktuell werden PQC-Algorithmen standardisiert, um der aufziehenden Gefahr für konventionelle asymmetrische Algorithmen durch Quantencomputer zu begegnen.
% Problem
Diese neuen Algorithmen müssen dann in bestehende Protokolle, Applikationen und Infrastrukturen eingebunden werden. Dabei ist mit Integrationsproblemen zu rechnen, die einerseits durch Inkompatibilitäten mit existierenden Standards und Implementierungen begründet sind, andererseits aber auch durch fehlendes Wissen der Softwareentwickler über die Handhabung von PQC-Algorithmen zustande kommen.
% Solution
Um Inkompatibilitäten beispielhaft aufzuzeigen, integrieren wir zwei unterschiedliche PQC-Algorithmen in zwei verschiedene bestehende Softwareprodukte (InboxPager E-Mail Client und TLS Implementierung der Bouncy Castle Bibliothek). Hierbei setzen wir auf die hoch-abstrahierende Krypto-Bibliothek eUCRITE, die Entwicklern das
Detailwissen über die korrekte Verwendung klassischer und PQC-Algorithmen abnimmt und damit bereits einige potentielle Implementierungsfehler
vermeidet. 
% Impact
Die dabei zutage getretenen Probleme bestätigen teilweise bereits bekannte Inkompatibilitäten, beinhalten aber auch neue, bisher nicht angesprochene Schwierigkeiten. 

\textbf{Stichworte:} Post-Quantum Verfahren, Kryptoagilität, eUCRITE-API, API  

\section{Einführung}
\label{sec:intro}

Quantencomputer sind Gegenstand der laufenden Forschung. Bei ausreichender Leistung, d.h. wenn Shor’s Algorithmus \autocite{Shor:1997:PAP:264393.264406}  auf einem Quantencomputer mit ausreichender Qubitlänge ausgeführt werden kann, wird man in der Lage sein, die derzeit verwendeten asymmetrischen Algorithmen wie RSA, DSA, ECDSA und ECDH zu brechen \autocite{chen_report_2016}. Der Bedarf an Post-Quanten-Kryptographie (PQC), insbesondere asymmetrischen Verfahren\footnote{Auch symmetrische Verfahren wie DES oder AES sind durch den Algorithmus von Grover (\cite{grover1997quantum}) bedroht, jedoch können
längere Schlüssel hier die Gefahr lindern.}, ist offensichtlich, 
da potentiell unsicher werdende asymmetrische Verfahren in vielen ausgerollten hybriden Kryptosystemen zu finden sind.

Dieser Beitrag stellt die im Rahmen des Forschungsprojektes \emph{Use-A-PQClib} \autocite{use-a-pqclib-site} entwickelte \eUCRITE{}-API \autocite{eUCRITEdocu} vor, die als Designziele eine gute Benutzbarkeit und Verständlichkeit für einen Entwickler sowie eine hohe Abstraktion von technischen Parametern wie beispielsweise Schlüssellängen von Krypto-Algorithmen aufweist.

Der Hauptteil beinhaltet darauf aufbauend einen Erfahrungsbericht bei der Integration der PQC-Verfahren McEliece \autocite{mceliece1978public} und SPHINCS+ \autocite{bernstein2015sphincs} auf Basis der \eUCRITE{}-API in zwei bestehende Softwareprodukte. Zum einen die Integration in InboxPager\footnote{\url{https://github.com/itprojects/InboxPager} (besucht am 29.12.2020)}, einen E-Mail Client für das Android Betriebssystem, zum anderen die Integration in die TLS Implementierung von Bouncy Castle\footnote{\url{https://www.bouncycastle.org} (besucht am 29.12.2020)}.
Bouncy Castle ist eine weit verbreitete Krypto-Bibliothek für Java und C\#, welche neben den grundlegenden Krypto-Operationen wie Verschlüsseln und Signieren auch eine TLS Implementierung zur Verfügung stellt. Hierbei konzentriert sich die Integration auf den Austausch der klassischen asymmetrischen Krypto-Verfahren gegen die oben genannten PQC-Verfahren der jeweiligen hybriden Kryptosysteme.
 
Damit soll aufgezeigt werden, mit welchen Herausforderungen und technischen Problemen Entwickler bei der Integration von PQC-Verfahren rechnen müssen. Des Weiteren werden die Vor- und Nachteile sowie die Implikationen einer hohen API-Abstraktion vorgestellt.

Abschließend wird ein Fazit gezogen und ein Ausblick auf weitere Schritte und noch zu adressierende Fragestellungen bei der Umstellung von klassischen auf PQC-Verfahren gegeben.

\section{Verwandte Arbeiten}

Der Dagstuhl Report \autocite{dagrep19451} berichtet über \textit{Biggest Failures in IT Security} und empfiehlt (unter anderem) Entwickler bei der Implementierung von Sicherheitsmechanismen stärker zu unterstützen. Dazu gehört zum einen  die Bereitstellung von Werkzeugen und Methoden, die es dem Entwickler leichter machen guten Code zu schreiben (\emph{to do the good/right thing}, Seite 20). Zum anderen gilt es, die verschiedenen Kenntnisse und Vorlieben der verschiedenen Entwickler- bzw. Benutzergruppen zu beachten.

\textcite{ott_identifying_2019} präsentieren zahlreiche Forschungsfragen zur Kryptoagilität \cite{Macaulay_Henderson_2019} und Migration nach PQC und decken dabei auf hohem Diskussionsniveau ein weites Feld an Themen ab, darunter Implementierungsaspekte. Dabei geht es den Autoren 
nicht nur um die Umsetzung der in mathematischen Formeln ausgedrückten PQC-Algorithmen auf verschiedensten Plattformen und in unterschiedlichen Programmiersprachen. Es ist ebenfalls von enormer Wichtigkeit, die implementierten Algorithmen so in vorhandene Systeme einzubringen, dass Kontinuität und Interoperabilität während der Migrationsphase erhalten bleiben.

Das Bundesamt für Sicherheit in der Informationstechnik (BSI) gibt Handlungsempfehlungen zur \textit{Migration zu Post-Quanten-Kryptografie} \autocite{bsi_pqc_2020} und empfiehlt (zumindest während der Migrationsphase) den Einsatz hybrider Lösungen, d.h. der Kombination klassischer und PQC-Algorithmen, und die entsprechende Anpassung kryptographischer Protokolle. Dabei soll die Umsetzung dem Prinzip der Kryptoagilität folgen, um auch zukünftige Empfehlungen und Standards entsprechend umsetzten zu können. Als besonders dringend gilt der Umstieg auf (hybride) PQC-Verfahren bei Schlüsseleinigungsverfahren zum Schutz langfristiger Geheimnisse.

\textcite{campagna_quantum_2015} nennen, neben der Beschreibung des State-of-the-Art in PQC und Anpassungsempfehlungen, um die Standards X.509, IKEv2, TLS 1.2, S/MIME und SSH2 PQC-bereit zu machen, wichtige Anwendungsfelder und -fälle für Kryptographie. Je nach bereits vorhandener Kryptoagilität beschränken sich die Empfehlungen auf die einfache Einführung neuer OIDs bzw. Cipher Suites oder beinhalten die mehr oder weniger tiefgreifende Anpassung vorhandener Standards. Die Gefahren von Quantencomputer-Angriffen auf typische Anwendungsfelder (Verschlüsselung, Authentisierung, ...) werden vorgestellt und in dieser Hinsicht besondere Verwundbarkeiten verschiedener Industriezweige diskutiert.

\textcite{crockett_prototyping_2019} stellen ihre Integration von PQC-Verfahren in die Protokolle TLS 1.2, TLS 1.3 und SSH2 vor, und berichten von den dabei zu bewältigenden Herausforderungen. Ein Beispiel für eine solche Herausforderung ist die zertifikatsbasierte Übermittlung mehrerer Schlüssel für unterschiedliche Verfahren, die in hybriden Umsetzungen benötigt werden. Ein anderes Beispiel sind Limitierungen für die Größe von Nachrichten zum Schlüsseltausch oder Signaturen, die teilweise implementierungsbedingt sind, aber teilweise auch auf  die Spezifikationen in den Standards zurückzuführen sind.  

\textcite{herath_mudiyanselage_next-generation_2019} untersucht, unter anderem, Eigenschaften von hybriden Signaturverfahren und deren Integration in verschiedene Implementierungen von X.509, TLS und S/MIME. Als wichtige Eigenschaften hybrider Signaturverfahren werden Unfälschbarkeit unter Signaturorakeln (EUF-CMA) und Nicht-Separierbarkeit\footnote{Eine hybride Signatur kann vom Angreifer nicht in eine Einzelsignatur umgeformt werden.} identifiziert. Die Integration eines zweiten Signaturschemas wurde über Erweiterungsmechanismen (X.509 extensions), nachgelagerte Authentifizierung (TLS) oder parallele/verschachtelte Signaturen (S/MIME) erreicht. In vielen Konstellationen führt die Einführung von Zusatzinformationen über ca. 40KiB zu Kompatibilitätsproblemen.

Die vorliegende Arbeit eruiert an zwei real umgesetzten PQC-Migrationen die dabei auftretenden Probleme und konkretisiert so die in den obigen Arbeiten aufgeführten theoretischen Überlegungen bzw. ergänzt die dort vorgestellten praktischen Herausforderungen.

\section{Verwendete Softwarekomponenten und Produkte}

Im Rahmen dieser Arbeit wurden Post-Quantum Verfahren in zwei Software-Produkte integriert. Hierbei wurde zum einen ein E-Mail-Client ausgewählt, da es sich hierbei mit über 300 Milliarden E-Mails pro Tag\footnote{\url{https://de.statista.com/statistik/daten/studie/252278/umfrage/prognose-zur-zahl-der-taeglich-versendeter-e-mails-weltweit/} (besucht am 29.12.2020)}  um eine der wesentlichen Anwendungen im Internet handelt. Zum anderen wurde mit TLS 1.2 ein Sicherheits-Protokoll gewählt, welches der aktuell gebräuchliche Standard zur Absicherung von HTTP im Internet ist.
Zur Bereitstellung der benötigten PQC-Algorithmen wurde auf die  Krypto-Bibliothek \eUCRITE{}-API zurückgegriffen. Da es sich dabei um eine Java-API handelt, sind die beiden betrachteten Implementierungen ebenfalls in Java geschrieben. Die API sowie die Softwareprodukte werden im Folgenden kurz vorgestellt.

\subsection{\eUCRITE{}-API}

\begin{figure}
    \centering
    \includegraphics[width=.7\textwidth]{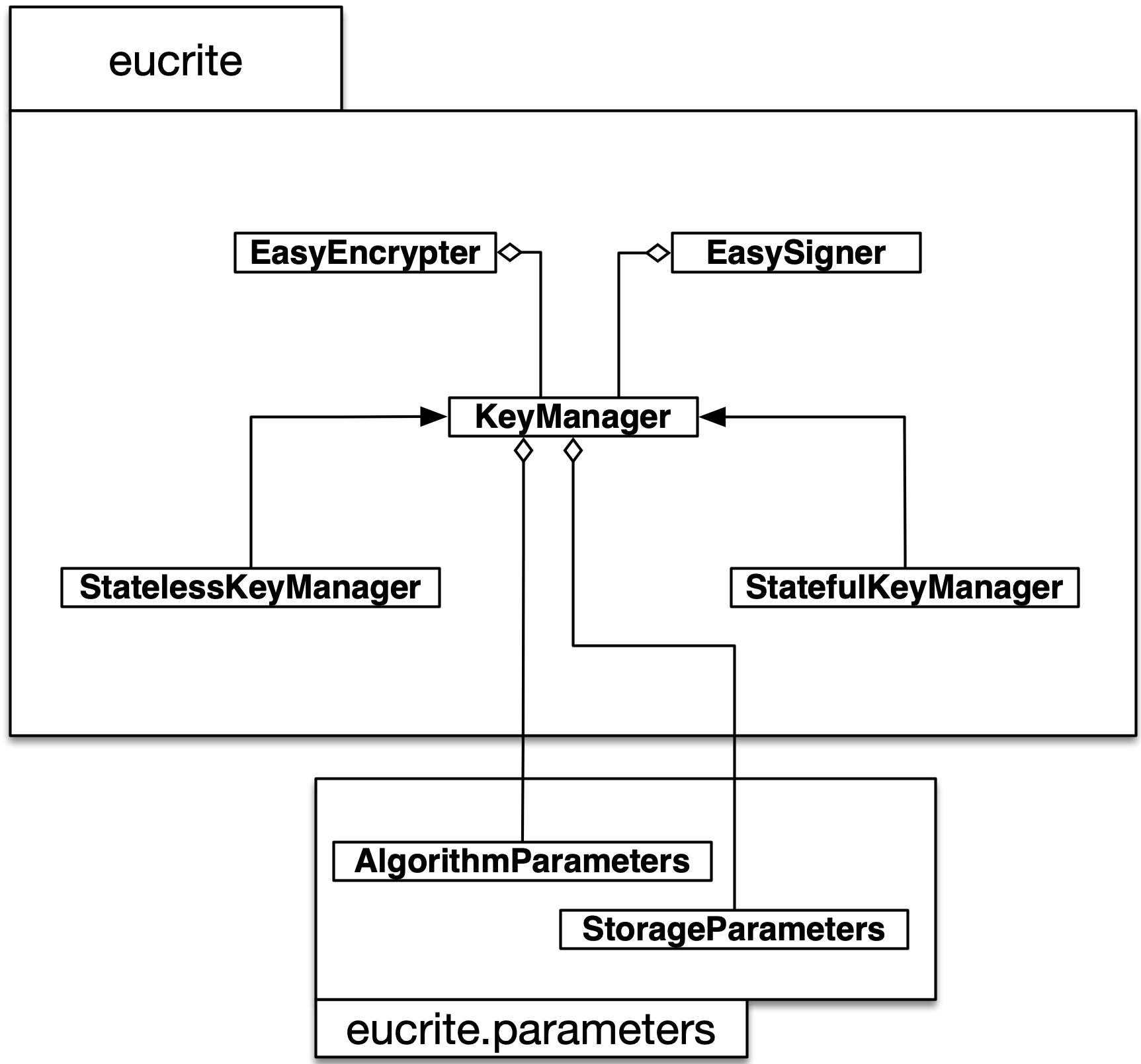}
    \caption{Klassendiagramm der \eUCRITE{}-API}
    \label{fig:eucrite_uml}
\end{figure}

Die \eUCRITE{}-API \cite{zeier2019api, eUCRITEdocu} ist eine kryptografische Bibliothek mit Fokus auf einfache Benutzbarkeit, die auch Laien die sichere Verwendung kryptografischer Verfahren ermöglicht. Dazu werden sogenannte \emph{Templates} eingesetzt, um die Auswahl der Algorithmen und Parameter für den Entwickler transparent zu gestalten. Die Auswahl wird durch die einfache Angabe des gewünschten Sicherheitsniveaus umgesetzt. Hier hat ein Entwickler die Wahl zwischen den Niveaus \texttt{LOW, MEDIUM, HIGH}. Das Sicherheitsniveau \texttt{HIGH} würde  intern dann beispielsweise zur Wahl der PQC Verfahren McEliece und SPHINCS+ führen.

Abbildung~\ref{fig:eucrite_uml} zeigt das Klassendiagramm der \eUCRITE{}-API. Die Klasse \texttt{EasyEncrypter} enthält Funktionen zur Ver- und Entschlüsselung, die Klasse \texttt{EasySigner} zur Signierung und Verifizierung von Daten. Beide Klassen nutzen einen \texttt{KeyManager} zur Speicherung des Schlüsselmaterials. Je nach Algorithmus wird ein \texttt{StatelessKeyManager} oder \texttt{StatefulKeyManager} verwendet. 
Bei der Erzeugung eines neuen Schlüsselpaares werden durch \texttt{AlgorithmParameters} der verwendete Algorithmus sowie dessen Parameter festgelegt. \texttt{StorageParameters} geben den Speicherort des Schlüsselmaterials an.

Die Usability der eUCRITE-API wurde in einer Reihe von vorgelagerten Nutzerstudien untersucht. Bei der zu diesem Zeitpunkt letzten Studie \cite{mci/Huesmann2020} wurde ein API-Usability Score nach Acar \cite{acar_comparing_2017} von 70,5 (aus maximal 100) erzielt. Im direkten Vergleich hat die API \textit{tink}\footnote{\url{https://github.com/google/tink} (besucht am 29.12.2020)} einen Score von 48,23 erreicht. In einer früheren Studie wurde insbesondere die Benutzbarkeit von zustandsbehafteten Verfahren in der \eUCRITE{}-API untersucht (\cite{zeier2019api}).

Das Codebeispiel in Listing~\ref{list:classicBC} zeigt -- vereinfacht -- Methodenaufrufe für die Signierung mit einem klassischen Verfahren (SHA256 und RSA) mithilfe  der Implementierung von Bouncy Castle, welcher als Provider über die Java Cryptography Extension (JCE) eingebunden wird.

Um auf ein PQC Verfahren zu wechseln, müsste hier die erste Zeile mit dem Codebeispiel aus Listing~\ref{list:pqcBC} ersetzt werden, die Zugriff auf PQC-Verfahren von Bouncy Castle erlaubt.
Beide Codebeispiele illustrieren, dass ein Entwickler eine Reihe von technischen
Parametern kennen und korrekt anwenden muss, z.B. den Namen und die Bitlänge des
zu verwendenden Hash-Algorithmus.

\begin{lstlisting}[caption={Signieren von Daten (klassisch) mit BouncyCastle},captionpos=b,label=list:classicBC]
Signature sig = Signature.getInstance("SHA256withRSA", "BC");
sig.initSign(privateKey);
byte[] toBeSigned = "Hallo Welt!".getBytes();
sig.update(toBeSigned, 0, toBeSigned.length);
byte[] signature = sig.sign();
\end{lstlisting}

\begin{lstlisting}[caption={Neue Zeile 1: Signieren von Daten (PQC) mit BouncyCastle},captionpos=b,label=list:pqcBC]
Signature sig = Signature.getInstance("SHA3-512WITHSPHINCS256", "BCPQC");
\end{lstlisting}

\begin{lstlisting}[caption={Signieren von Daten mit der \eUCRITE{}-API},captionpos=b,label=list:pqcEucrite]
AlgorithmParameters algorithmParameters = 
     AlgorithmParameters.Template.Signature.Security_Level.HIGH.getParameters(); 
KeystoreParameters keystoreParameters = new KeystoreParameters(keyStoreFile, "password"); 
EasySigner signer = EasySigner.withNewKey(algorithmParameters, keystoreParameters); 
byte[] signature = signer.sign("Hallo Welt!");
\end{lstlisting}

In eUCRITE kann hingegen das bereits erwähnte Template verwendet werden (siehe Codebeispiel in Listing~\ref{list:pqcEucrite}). Die Wahl des Sicherheitslevels \texttt{HIGH} für Signaturen führt intern z.\ Zt. zur Auswahl von SHA3-512 und SPHINCS+. 

In einem späteren Schritt soll \eUCRITE{} durch einen Update Mechanismus stets ein zur Laufzeit als sicher geltendes Verfahren ausgewählen, d.h. sollte SPHINCS+ in Zukunft gebrochen werden, würde hier ein anderer Algorithmus zum Einsatz kommen.

\subsection{InboxPager (Android E-Mail Client)}

Als erster Untersuchungsgegenstand zur Integration von PQC-Verfahren auf Basis der \eUCRITE{}-API wurde ein E-Mail Client ausgewählt, da hier sowohl asymmetrische Verschlüsselung als auch Signaturen zum Einsatz kommen. Die Auswahl des Clients erfolgte auf Basis der folgende Kriterien: 
\begin{enumerate}
    \item Klassische Kryptographie-Verfahren müssen bereits integriert sein
    \item die kryptographischen Funktionen sind gut getrennt vom restlichen Code
    \item die Anwendung ist nicht zu umfangreich
    \item die Anwendung ist in Java geschrieben.
\end{enumerate}

\begin{figure}
    \centering
    \includegraphics[width=0.8\textwidth]{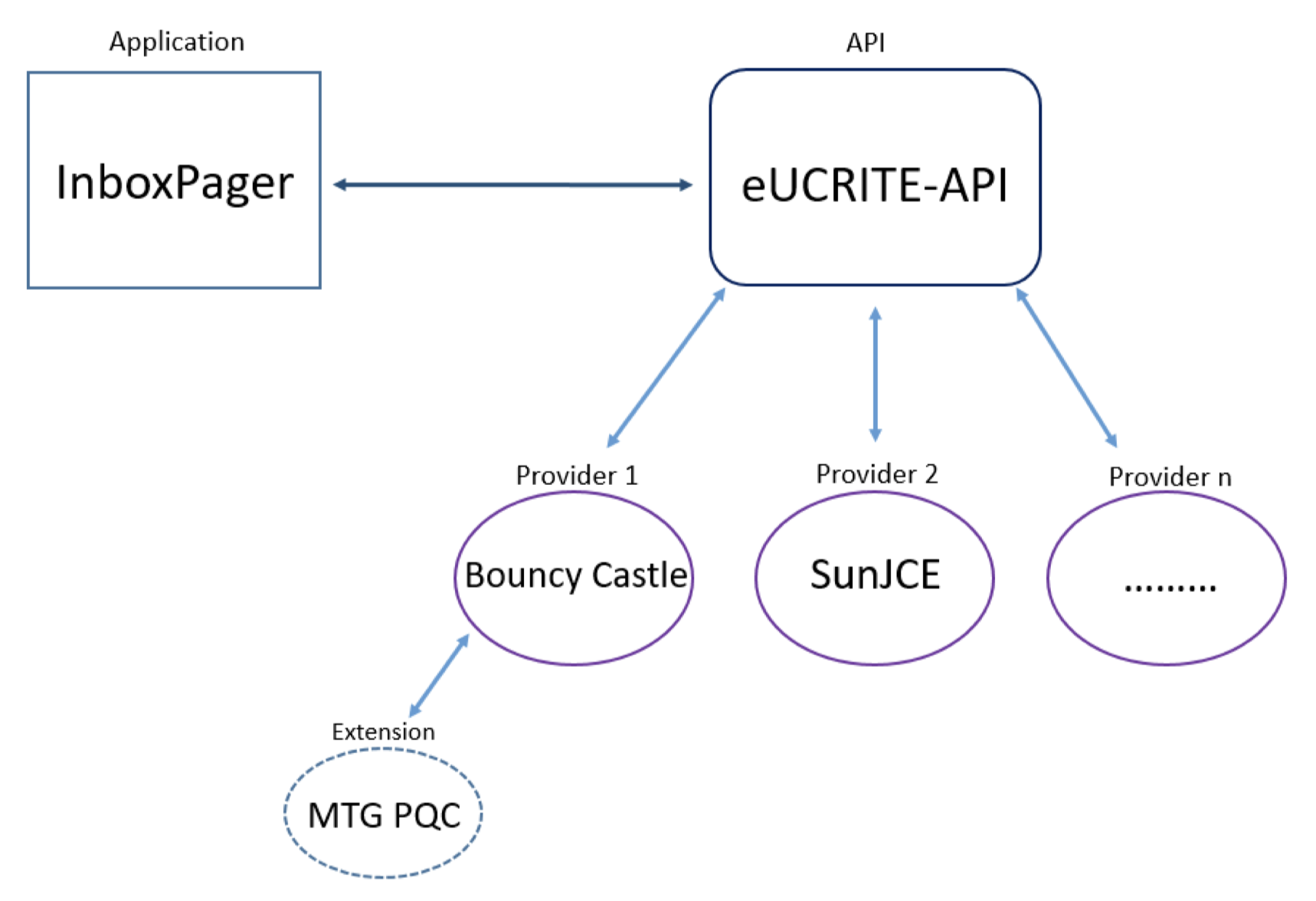}
    \caption{Abhängigkeiten der kryptografischen Bibliotheken \autocite{Yousofzai_Integration_2020}}
    \label{fig:java_crypto_architecture}
\end{figure}

Durch eine erste Recherche standen die folgenden drei E-Mail Clients zur Auswahl: K9 \cite{k9}, InboxPager \cite{InboxPaper} und FairMail \cite{FairMail}. Anhand der oben genannten Kriterien fiel die Wahl auf InboxPager (Version 4.5).

Abbildung~\ref{fig:java_crypto_architecture} zeigt die Abhängigkeiten der verwendeten kryptografischen Bibliotheken. In der überarbeiteten InboxPager Implementierung werden die Funktionen zur Verschlüsselung und Signierung der E-Mails über die \eUCRITE{}-API implementiert. Intern greift diese, abhängig vom Krypto-Verfahren, auf verschiedene Provider zurück. Die beiden PQC Verfahren Classic McEliece und SPHINCS+ werden über die MTG PQC Extension\footnote{Siehe Abschnitt \ref{sub:technA}} des Bouncy Castle Providers angesprochen.

Für weitere Details wird auf die Bachelorarbeit von \textcite{Yousofzai_Integration_2020} verwiesen, in deren Rahmen wesentliche Teile der Integration entstanden. 

\subsection{TLS Implementierung in Bouncy Castle}

Aufgrund der zum Durchführungszeitpunkt immer noch sehr weiten Verbreitung von TLS 1.2 (vgl. \autocite{crockett_prototyping_2019}) sowie der Verfügbarkeit einer TLS 1.2 Implementierung in Bouncy Castle (Version 1.66) wurde die Integration der \eUCRITE{}-API in diese TLS Implementierung untersucht.
Wesentliche Teile der Integration sind im Rahmen der Bachelorarbeit von \textcite{Merz_Integration_2020} entstanden.

\begin{lstlisting}[caption={Verwendete Cipher Suite Definition}, label=list:def]
public static final int TLS_CME_SPX_WITH_AES_25_CBC_SHA512 = 0x1306;
\end{lstlisting}

Wie in Abschnitt \ref{sec:intro} erwähnt, lag der Fokus dieser Arbeit auf dem 
Austausch der asymmetrischen Krypto-Verfahren in einem hybriden Kryposystem wie 
TLS. Da zum aktuellen Zeitpunkt seitens der IANA noch keine passende TLS Cipher Suite mit PQC-Verfahren definiert ist \cite{tls-iana}, wurde für die Implementierung der Wert \texttt{0x1306} verwendet, der noch nicht von der IANA vergeben wurde\footnote{Status \emph{unassigned}} (siehe Listing \ref{list:def}).

\begin{figure}
    \centering
    \includegraphics[width=0.8\textwidth]{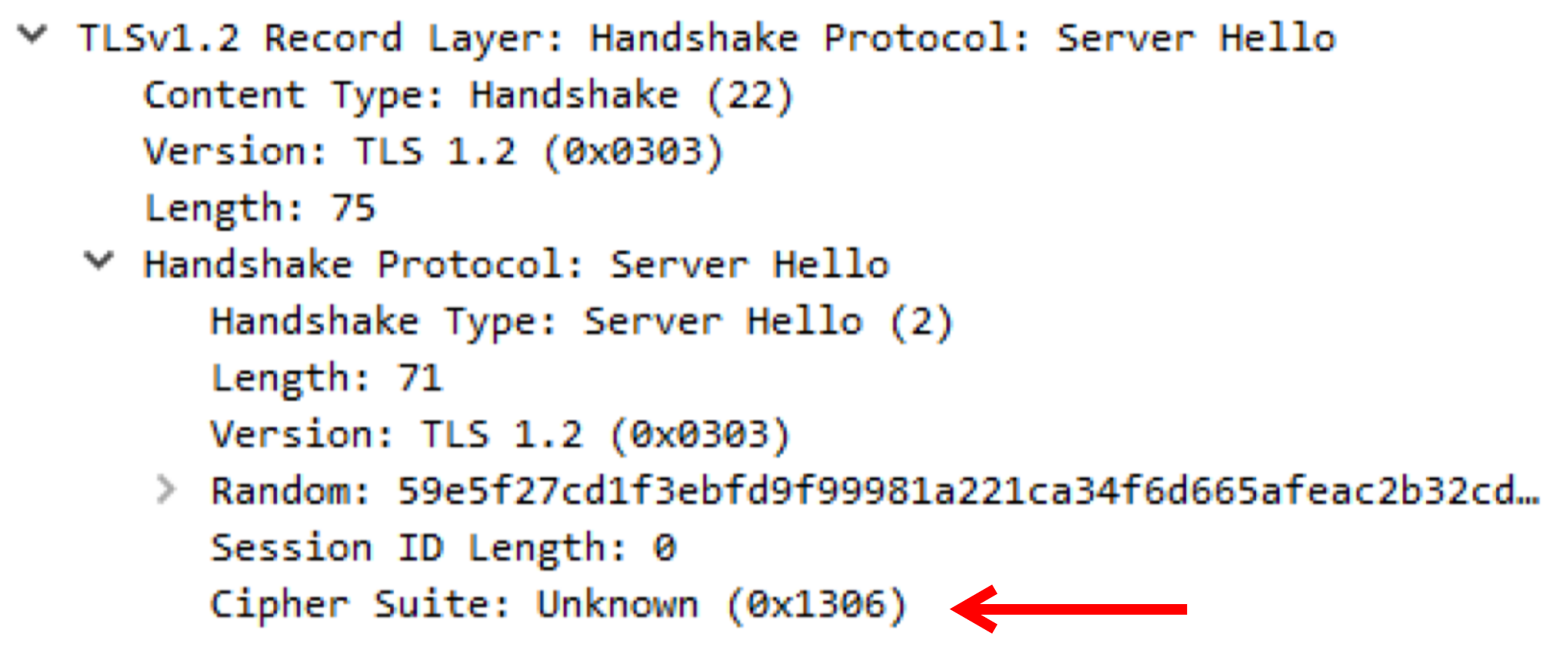}
    \caption{TLS \texttt{Server Hello} mit neuer, für \textsc{Wireshark} unbekannter Cipher Suite \autocite{Merz_Integration_2020}}
    \label{fig:wireshark_ciphersuite}
\end{figure}

Der PQC-basierte TLS Handshake lief erfolgreich zwischen zwei Knoten ab, die
beide die neue Cipher Suite verarbeiten konnten. Die Analyse des Datenverkehrs mithilfe von \textsc{Wireshark} meldet erwartungsgemäß an dieser Stelle eine unbekannte Cipher Suite (siehe Abbildung~\ref{fig:wireshark_ciphersuite}).

\section{Erkenntnisse}

Im Folgenden wird auf die wesentlichen Hürden eingegangen, die bei der Integration der PQC-Verfahren in hybride Krytosysteme auftraten. Diese lassen sich in konzeptionelle/organisatorische (Abschnitt~\ref{sub:concepts}) und technische Aspekte (Abschnitt~\ref{sub:technA}) unterscheiden.

\subsection{Konzeptionelle/organisatorische Aspekte}
\label{sub:concepts}

Die \eUCRITE{}-API bietet bei der Wahl der Krypto-Verfahren eine höhere Abstraktion als herkömmliche Bibliotheken. Ein Entwickler spezifiziert zur Compile-Zeit nur noch sein gewünschtes Sicherheitslevel (\texttt{LOW, MEDIUM, HIGH}). Intern wählt \eUCRITE{} dann die passenden Krypto-Verfahren aus. Beispielsweise könnte für \texttt{LOW} noch RSA ausreichend sein, während \texttt{HIGH} bereits auf das Quantencomputer-resistente McEliece-Verfahren zurückgreift. Diese genau spezifizierte Angabe zur Wahl des Krypto-Verfahrens wird dann z.B. im TLS-Handshake verwendet. Hier offenbart sich eine semantische Unschärfe, sollten \eUCRITE{}-basierte Kommunikationspartner zu unterschiedlichen Compile-Zeiten unterschiedliche Verfahren für ein Sicherheitslevel ausgewählt haben. Ein Sicherheitslevel \texttt{HIGH} im Jahr 2021 könnte zur Wahl eines anderes Verfahrens führen als ein Sicherheitslevel \texttt{HIGH} im Jahr 2031. Es muss also eine Möglichkeit geschaffen werden (z.B. über einen Update Mechanismus der \eUCRITE{}-API oder durch Versionsnummern der \eUCRITE{}-API) dies zu erkennen und programmtechnisch zu behandeln.

Zum jetzigen Zeitpunkt fehlen Spezifikationen für Cipher Suites, die PQC-Verfahren unterstützen. Es gibt erste Ansätze\footnote{z.B. \url{https://tools.ietf.org/html/draft-campagna-tls-bike-sike-hybrid-01} (besucht am 29.12.2020)} , die jedoch nicht alle Kombinationen von Verfahren abdecken. So haben wir für unsere Umsetzung TLS\_CME\_SPX\_WITH\_AES\_256\_CBC\_SHA512 gewählt. In Folge einer Standardisierung könnte dieser Wert ungültig werden.

\subsection{Technische Aspekte}
\label{sub:technA}

Die Analyse des Netzwerkverkehrs unserer prototypischen Implementierungen zeigt, dass bei unserem TLS Handshake, basierend auf McEliece, ca. 1.5 MB an Nutzdaten bei der Übermittlung des Zertifikats anfallen. Diese Beobachtung deckt sich mit anderen Arbeiten \autocite{sikeridis_post-quantum_2020, kampanakis_viability_2018, burstinghaus-steinbach_post-quantum_2020} und könnte sich negativ auf die User-Experience beim Surfen im Internet auswirken, da der TLS Handshake zu viel Zeit konsumiert, insbesondere bei sog. \textit{lossy networks} verschärft sich dieses Problem \autocite{paquin_benchmarking_2019}.

Unsere Wahl der Programmiersprache Java für die \eUCRITE{}-API und die E-Mail Anwendung InboxPager sowie die Bibliothek Bouncy Castle zog die folgenden technischen Anpassungen nach sich: Die Implementierung der PQC-Verfahren wurde uns in Form einer Erweiterung der Bouncy-Castle-Bibliothek durch unseren Projektpartner, der MTG AG aus Darmstadt \cite{MTG}, bereitgestellt, die aus Gründen der Performance über das Java Native Interface die eigentlichen Algorithmen nah an der Ziel-Hardware in der Programmiersprache C umsetzt und auf Prozessorarchitektur-Optimierungen setzt. Dies limitiert die Portabilität unserer Lösungen.

Weiter ist es aufgrund der Größe des McEliece Schlüsselmaterials notwendig, die Größe des JVM Thread Stack zu erhöhen, da das Schlüsselmaterial über den Stack an die native C-Anbindung (via Java Native Interface) übergeben wird.

Schließlich sei erwähnt, dass das Android OS bereits mit einer Bouncy-Castle-Bibliothek als Java Cryptographic Service Provider ausgeliefert wird. Dieser Provider muss zunächst deaktiviert werden, da es ansonsten zur Auswahl des falschen Providers kommt. Weiter wird Google für Android in Zukunft obligatorisch auf Conscrypt\footnote{\url{https://conscrypt.org} (besucht am 29.12.2020)}  als Implementierung für die klassischen Krypto-Algorithmen setzen, was für die Integration von PQC-Verfahren Anpassungen beim Zusammenspiel mehrerer Provider erfordern wird.

\section{Fazit und Ausblick}

\subsection{Fazit}

Die vorliegende Arbeit berichtet über praktische Erkenntnisse bei der Integration von PQC-Verfahren in bestehende Softwareprodukte. Die Integration erfordert zum Teil detaillierte Kenntnisse bzw. Änderungen an der Laufzeitumgebung, wie der beobachtete Thread-Stack-Fehler zeigt. Um die Interoperabilität zwischen (beliebigen) Anwendungen sicherstellen zu können, bedarf es weiterer Standardisierung, z.B. bezüglich der zu verwendenden Konstanten für Cipher Suites. Der Punkt Endbenutzerfreundlichkeit bedarf ebenfalls der Aufmerksamkeit. So müssen z.B. die PQC-Algorithmen hardwarenah ausgeführt werden, um eine akzeptable Wartezeit zu erreichen. Obwohl die Integration von PQC in TLS-1.2 technisch erfolgreich war, ist es anzuraten hier auf modernere Protokolle (z.B. TLS-1.3 oder Google QUIC) zu setzen, da diese den Overhead zum Aufbau einer Sitzung, und damit die Ausführung von PQC-Verfahren deutlich reduzieren werden.

\subsection{Ausblick}

Weiterführend soll das korrekte Zusammenwirken der hier vorgestellten Abstraktionsmechanismen über System-, und Zeitgrenzen untersucht werden. Hierfür wird, insbesondere für die zeitliche Komponente, der erwähnte Update-Mechanismus von Bedeutung sein. Neben der Durchführung von weiteren Integrationen und Tests sollen die erzielten Ergebnisse in das Design und die Funktionalität der \eUCRITE{}-API zurückfließen.

\subsection*{Danksagung}

Dieser Beitrag wurde im Rahmen der Innovationsförderung des Landes Hessen aus Mitteln der LOEWE – Landes-Offensive zur Entwicklung Wissenschaftlich-ökonomischer Exzellenz, Förderlinie 3: KMU-Verbundvorhaben unter HA-Projekt-Nr.: 633/18-56 gefördert.

\printbibliography 

\typeout{get arXiv to do 4 passes: Label(s) may have changed. Rerun}
\end{document}